\pdfoutput=1 
\documentclass[aps,prx,twocolumn, superscriptaddress,amsfont,graphicx,nofootinbib,preprintnumbers,10pt,UTF8，hyperref]{revtex4-1}
\usepackage{amsmath}
\usepackage{amssymb}
\usepackage{color}
\usepackage{xcolor}
\usepackage{graphicx}
\usepackage{nameref}
\usepackage{textcomp}
\usepackage{xspace,slashed}
\usepackage[utf8]{inputenc}
\usepackage{subcaption}
\usepackage{multirow}
\usepackage{nccmath} 
\usepackage{slashed}
\usepackage{orcidlink}
\usepackage{booktabs}
\usepackage{listings} %
\usepackage{hyperref}

\begin{document}

\title{
Explainable AI-assisted Optimization for Feynman Integral Reduction
}

\author{Zhuo-Yang Song~\orcidlink{0009-0001-9727-7908}}
\email{zhuoyangsong@stu.pku.edu.cn}
\affiliation{School of Physics, Peking University, Beijing 100871, China}
\author{Tong-Zhi Yang~\orcidlink{0000-0001-5003-5517}}
\email{tongzhi.yang@physik.uzh.ch}
\affiliation{Physik-Institut, Universit\"at Z\"urich, Winterthurerstrasse 190, 8057 Z\"urich, Switzerland} 
\author{Qing-Hong Cao}
\email{qinghongcao@pku.edu.cn}
\affiliation{School of Physics, Peking University, Beijing 100871, China}
\affiliation{Center for High Energy Physics, Peking University, Beijing 100871, China}

\author{Ming-xing Luo}
\email{mingxingluo@csrc.ac.cn}
\affiliation{Beijing Computational Science Research Center, Beijing 100193, China}

\author{Hua Xing Zhu}
\email{zhuhx@pku.edu.cn}
\affiliation{School of Physics, Peking University, Beijing 100871, China}
\affiliation{Center for High Energy Physics, Peking University, Beijing 100871, China}

\preprint{ZU-TH 07/25}

\begin{abstract}

We present a novel approach to optimizing the reduction of Feynman integrals using integration-by-parts identities. By developing a priority function through the FunSearch algorithm, which combines large language models and genetic algorithms, we achieve significant improvements in memory usage and computational efficiency compared to traditional methods. Our approach demonstrates substantial reductions in the required seeding integrals, making previously intractable integrals more manageable. Tested on a variety of Feynman integrals, including one-loop and multi-loop cases with planar and non-planar configurations, our method demonstrates remarkable scalability and adaptability. For reductions of certain Feynman integrals with many dots and numerators, we observed an improvement by a factor of 3058 compared to traditional methods. This work provides a powerful and interpretable framework for optimizing IBP reductions, paving the way for more efficient and practical calculations in high-energy physics.
\end{abstract}

\maketitle



\section{Introduction}
\label{sec:intro}
Precision theoretical predictions in particle physics are crucial for rigorously testing the Standard Model and interpreting experimental results, as even minor discrepancies between theory and observation can reveal new physics. Due to the asymptotic freedom of Quantum Chromodynamics (QCD) and collinear factorization, theoretical predictions for hadronic observables at high-energy hadron colliders can be formulated within the framework of perturbative theory. In perturbative quantum field theory (QFT), two primary methods are commonly used to compute the Feynman integrals that contribute to scattering amplitudes or cross-sections: integration-by-parts (IBP) reductions~\cite{Chetyrkin:1981qh,Tkachov:1981wb} and the differential equation (DE) method~\cite{Kotikov:1990kg,Gehrmann:1999as,Henn:2013pwa}. Notably, the DE method itself relies heavily on IBP reductions, making IBP reductions a major computational bottleneck in high-precision calculations. 

Traditionally, most of the IBP reductions were carried out using the Laporta algorithm and Laporta seeding~\cite{Laporta:2000dsw}, which, while effective, become increasingly inefficient as the complexity of the problem grows. In recent years, the finite field methods~\cite{vonManteuffel:2014ixa, Peraro:2016wsq}, which avoid intermediate expression swell, have been introduced to enhance IBP reductions and improve computational feasibility. However, these approaches also encounter challenges when applied to highly intricate problems. Additionally, alternative approaches such as syzygy techniques~\cite{Gluza:2010ws,Larsen:2015ped,Bohm:2017qme,Lee:2014tja,Bitoun:2017nre,Agarwal:2020dye}, intersection number theory~\cite{Mastrolia:2018uzb,Frellesvig:2019uqt}, and block-triangular form improved reduction~\cite{Liu:2018dmc,Guan:2019bcx,Guan:2024byi} have emerged, offering new perspectives on IBP reductions. Building on these techniques, several public packages have been developed to perform IBP reductions, including:~\texttt{Air}~\cite{Anastasiou:2004vj}, ~\texttt{LiteRed}~\cite{Lee:2012cn,Lee:2013mka}, \texttt{FIRE6}~\cite{Smirnov:2019qkx,Smirnov:2023yhb}, \texttt{Reduze}~\cite{vonManteuffel:2012np,Studerus:2009ye}, \texttt{Kira}~\cite{Maierhofer:2017gsa,Klappert:2019emp,Klappert:2020nbg}, ~\texttt{Forcer}~\cite{Ruijl:2017cxj}, \texttt{FiniteFlow}~\cite{Peraro:2019svx}, \texttt{NeatIBP}~\cite{Wu:2023upw}, \texttt{Blade}~\cite{Guan:2024byi},  \texttt{AmpRed}~\cite{Chen:2024xwt}.

To further enhance the efficiency of IBP reductions, we introduce a novel approach that optimizes the reduction of Feynman integrals through a priority function developed via the FunSearch algorithm~\cite{funsearch}. FunSearch combines large language models (LLMs) with genetic algorithms to evolve and refine priority functions that can minimize IBP system size and improve computational efficiency. By leveraging the scalability and interpretability of LLMs, this approach provides heuristic solutions that can be extended to more complex problems, offering a new avenue for advancing precision calculations in particle physics.

Our work focuses on optimizing the IBP reductions process by identifying an optimal subset of seeding integrals that are sufficient to solve the target integrals. We demonstrate the effectiveness of our approach through explicit examples involving one-loop and multi-loop Feynman integrals, including both planar and non-planar six-particle phase-space integral families. Our results show that the priority function method achieves substantial improvements in memory usage and computational efficiency compared to traditional methods, enabling the reduction of previously intractable integrals.

In recent years, we have witnessed increasing applications of Artificial Intelligence (AI) in high-energy theoretical physics, despite the inherent tension between the need for interpretability in theoretical physics and the probabilistic, black-box nature of AI systems. Substantial progress has been made, including applications of machine learning to the string-theory landscape~\cite{He:2024gnk,He2023,He:2019vsj,He:2020mgx}, scattering amplitudes~\cite{Cai:2024znx,Dixon:2022rse,Schwartz:2021ftp,Dersy:2022bym,Dersy:2023job,Cheung:2024svk,Demirtas:2023fir,Bhattacharya:2024chz}, jet physics~\cite{Komiske:2016rsd,Komiske:2017ubm,Heimel:2018mkt,Qu:2019gqs,Gong:2022lye,Bright-Thonney:2023gdl,Metodiev:2023izu,Desai:2024yft,Gambhir:2024dtf,Heimel:2023mvw,Butter:2022rso}, Parton Distribution Functions~\cite{Forte:2002fg,NNPDF:2014otw,Liu:2022plj}, as well as advances in understanding neural networks through the framework of quantum field theories~\cite{Badger:2020uow,Aylett-Bullock:2021hmo,Fedkevych:2022mid,Calisto:2023vmm, halverson2019computational,gukov2021learning,halverson2021neural,halverson2019branes,JHalverson}. While most previous studies have focused on learning underlying structures from theoretical data, our work takes a different approach. We introduce the priority function method for IBP reduction in Feynman integral calculations, where rather than learning from data, we formulate a clearly defined theoretical target for AI optimization in a fully explainable way.

We have achieved substantial improvements in both computational efficiency and memory usage through our priority function method, clearly demonstrating the usefulness of AI in cutting-edge analytic theoretical research. These enhancements are critical for making multi-loop calculations more feasible and scalable, thereby enabling researchers to explore more complex and higher-order processes that were previously computationally prohibitive. By integrating the scalability and interpretability of LLMs with the optimization capabilities of genetic algorithms, our FunSearch approach not only addresses existing computational bottlenecks but also sets a new benchmark for future research with AI assistance. 

In the following, we present some necessary concepts for Feynman integrals and FunSearch. Subsequently, we employ FunSearch algorithm to search for the best-estimated priority function for one-loop bubble integrals, which helps to greatly reduce the size of IBP system. The obtained priority function is then directly generalized to the integrals with any number of loops and legs. Finally, we show several explicit examples where the priority function is used to efficiently perform IBP reductions for six-particle phase-space integrals in both planar and non-planar cases.

\section{Overview of Feynman integral reduction}
In this section, we provide an overview of the Feynman integral and the traditional methods to perform IBP reductions. This background will set the stage for introducing our approach in the subsequent sections.

\subsection{Feynman integral definition}
A family of $j$-loop Feynman integrals is defined as follows (see for example~\cite{Smirnov2012}):
\begin{align}
    I(n_1,n_2,&\cdots, n_m) = \int \left(\prod_{i=1}^j \frac{d^D l_i}{i \pi^{d/2-1}} \right) i(n_1,n_2,\cdots,n_m) \nonumber \\
    & = \int \left(\prod_{i=1}^j  \frac{d^d l_i}{i \pi^{d/2-1}} \right) \frac{1}{D_1^{n_1} \,D_2^{n_2}\,\cdots D_m^{n_m}}\,,
\end{align} 
where $D$ is the spacetime dimension, the propagators $D_m$ depend on loop momentum $l_j$ and external momentum $p_1,p_2,\cdots, p_E$. The propagator's powers $n_m$ are positive or negative integers. The propagators which allow negative powers only are called irreducible scalar products (ISPs). The integrals are classified into \textit{sectors}
$   (\theta_1, \theta_2, \cdots, \theta_m)$ with
$\theta_i = \Theta(n_i-1/2)$. It's convenient to define several parameters: the number of different denominators $t=\sum_i \theta_i$, the total power of denominators $r = \sum_i \theta_i n_i$, the total power of numerators $s = \sum_i (\theta_i-1) n_i$, as well as the dots $d = r- t$. The integrals of total derivative are zero in dimensional regularization, this gives the IBP identities: 
\begin{align}
\label{eq:IBP}
    0= \int \left(\prod_{i=1}^j \frac{d^d l_i}{i \pi^{d/2-1}} \right) \frac{\partial }{\partial l_i^{\mu}} \bigg( q_k^\mu \, i(n_1,n_2,\cdots, n_m) \bigg) \,,
\end{align}
where $q = l_1,\cdots l_j, p_1, \cdots p_E$. These identities are a set of recurrence relations which relate a large amount of Feynman integrals to a basis of integrals, called \textit{master integrals}. In the presence of cut propagators, the reverse unitarity method~\cite{Anastasiou:2002yz} is introduced to derive IBP identities,
\begin{align}
\delta(D_i) = \left[\frac{1}{D_i}\right]_\text{cut} = \frac{1}{2 i \pi } \left(  \frac{1}{D_i -i 0 } - \frac{1}{D_i + i 0} \right)\,.
\end{align}

\subsection{Laporta seeding and improved seeding}

A standard approach for solving IBP reduction problems is the Laporta algorithm~\cite{Laporta:2000dsw}. This method works by substituting specific integer values for the propagator indices in the symbolic IBP identities in Eq.~\eqref{eq:IBP}, then solving the resulting large-scale linear system to express generic Feynman integrals in terms of master integrals. The specific integrals generated through this index assignment process are referred to as \textit{seeding integrals}.

To systematically generate these seeding integrals, one typically employs predefined patterns. The conventional Laporta seeding scheme follows a bounding strategy $s\leq s_{\text{max}},\, d \leq d_\text{max}$. However, this approach often produces a large number of redundant equations that do not contribute to new constraints, significantly increasing the computational complexity of solving the large linear system. Recent advances in the field have introduced more sophisticated seeding strategies~\cite{Driesse:2024xad,
Guan:2024byi,Bern:2024adl} that dramatically reduce this redundancy while maintaining system completeness. These seeding strategies, referred to as \textit{improved seeding}, propose that, in addition to Laporta seeding, $s$ can be reduced by $n+f$ for sectors missing $n$ propagators relative to the primary sector, where $f$ is an adjustable parameter.

Building on numerous past efforts to enhance the efficiency of Feynman integral reduction, including the development of improved seeding methods, we propose a novel advancement in IBP reduction. Our approach leverages machine learning techniques, with a particular focus on the FunSearch approach~\cite{funsearch}.

\section{Overview of FunSearch}

With the background on Feynman integral reduction established, we now turn to our approach for optimizing this process using the FunSearch algorithm~\cite{funsearch}. We propose a priority function that allows FunSearch to combine LLMs with genetic algorithms to discover efficient algorithms that can significantly reduce the number of required seeding integrals.

FunSearch is an innovative approach that combines LLMs with genetic algorithms to discover novel solutions to complex mathematical problems. This method is particularly suitable for problems that are "easy to evaluate but difficult to solve" such as combinatorial optimization and certain mathematical problems~\cite{funsearch}. FunSearch operates by evolving programs that describe how to solve the problem, rather than directly searching for the solution itself. This approach endows the discovered solutions with scalability and interpretability.

The scalability allows FunSearch to repeatedly run on simpler problems to obtain efficient and heuristic solutions. The interpretability enables humans to further generalize and improve upon these heuristic solutions, thus making a machine-expert collaborative workflow possible. On the other hand, extensive searches using FunSearch can avoid the selective use of instances favorable to human experts' solutions, thereby providing more generalizable results. This leads to more objective and comprehensive solutions in the field of high-energy physics.

The objective of FunSearch in this work is to optimize the seeding in the IBP reductions process by developing a priority function that sorts seeding integrals, thereby accelerating IBP reductions. The detailed implementation is presented in Sec.~\ref{sec:encoding}.

\begin{widetext}

    \begin{figure}[htp]
    \centering
    \includegraphics[width=1.0\textwidth]{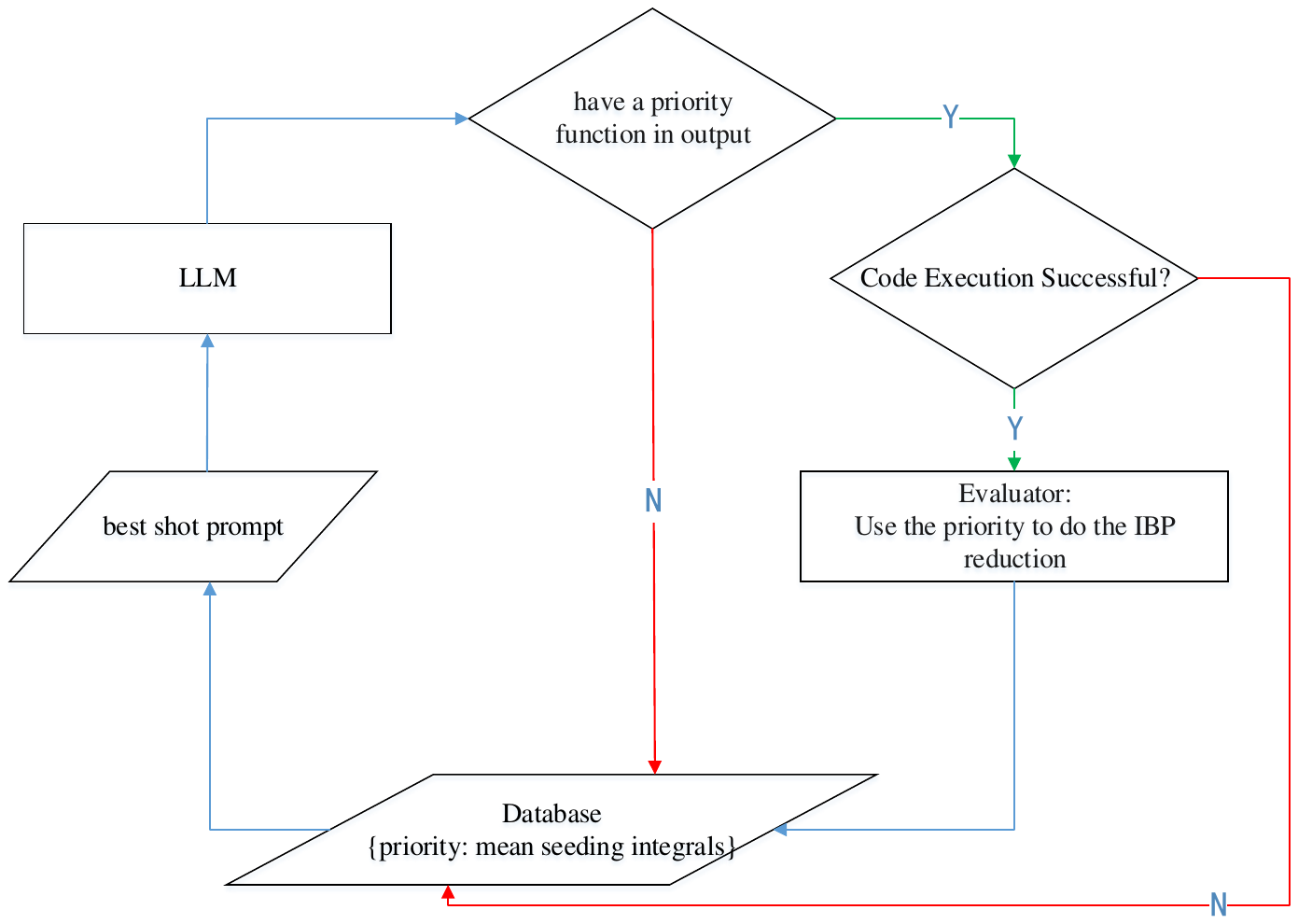}
    \caption{This flowchart outlines the operational sequence of the FunSearch algorithm, which is designed to optimize priority functions for IBP reduction tasks. The process begins with retrieving of one or two existing priority functions from the database. These functions, coupled with a concise explanation of the IBP reduction process, are formatted into a best-shot prompt. This prompt serves as input for a pre-trained LLM, which generates new priority functions based on the provided information. FunSearch then extracts these newly generated priority functions from the LLM's output.
    The extracted priority functions undergo a validation process to determine if they are executable. If successful, the evaluator utilizes these functions to perform IBP reductions, assessing their effectiveness. Functions that demonstrate superior performance are subsequently integrated back into the database, enriching it for future iterations of the algorithm. This cyclical process ensures continuous improvement and refinement of the priority functions used in IBP reductions.}
    \label{fig:flow chart}
    \end{figure}
\end{widetext}

FunSearch consists of two key components: a pre-trained large LLM and an evaluator that assesses the quality of priority functions through scores. FunSearch evolves an initially low-scoring priority function into a higher-scoring one through iterative operations of these components. The iteration flow chart is shown in Fig.~\ref{fig:flow chart}.

 During the iterative process, we employ an island-based evolutionary approach to encourage exploration and prevent the functions from falling into local optima. The programs evolve independently on different "islands," and periodically, the worst-performing island is reset with new programs from the best-performing island~\cite{funsearch}.

Compared to traditional search methods, FunSearch has several advantages:

\textbf{Scalability:} By searching in the space of programs, FunSearch can handle problems with large search spaces more efficiently than direct solution search methods. In the context of IBP reductions, the number of required seeding integrals is often much smaller than the total number of initial seeding integrals. This suggests that optimal seeding may depend on the specific target integral. However, designing a separate IBP reduction algorithm for each target integral is impractical, so common methods like Laporta seeding~\cite{Laporta:2000dsw} are typically only minimally dependent on the target integral. This results in a large number of redundant seeding integrals. FunSearch, with its scalability, has the potential to resolve this contradiction.

\textbf{Interpretability:} The programs generated by FunSearch are typically interpretable, facilitating insights and further refinement by domain experts. This means that FunSearch can first be run on a relatively simple IBP reductions, and then domain experts can extend the heuristic functions generated by FunSearch to more complex IBP reductions. This can significantly enhance the efficiency of FunSearch, thereby expanding its application scope and reducing deployment costs.

\textbf{Robustness:} Combining the creativity of LLMs and evolutionary exploration enables FunSearch to avoid local optima and discover novel solutions. In traditional IBP reduction methods, human experts often select specific seedings based on their experience and preferences. In contrast, FunSearch uses an automated approach that eliminates such selectivity, ensuring objectivity and generality in the search process.

In summary, FunSearch provides a powerful framework that leverages the strengths of LLMs and evolutionary algorithms to solve complex and innovative problems. Its ability to generate and refine programs offers a scalable and interpretable problem-solving approach, making it a valuable tool in high-energy physics, especially when innovative solutions are required. 

Recent work has demonstrated that FunSearch can be further optimized in multiple aspects~\cite{HSEvo,EoH,ReEvo,UBER,MCTS}, enabling it to tackle more complex problems while showcasing its flexibility in solving problems across various domains~\cite{funapply,funsearchbenchmark}.

\section{Encoding IBP reductions through priority function}
\label{sec:encoding}

With this background in mind, we will now discuss how to apply FunSearch to IBP reductions.

In high-energy physics, the Laporta algorithm~\cite{Laporta:2000dsw} is widely used to solve the IBP reduction problem for Feynman integrals. Previous optimization methods over Laporta seeding, such as those in~\cite{Driesse:2024xad,Guan:2024byi,Bern:2024adl}, have imposed constraints on the search space that may not be optimal. To optimize this process more effectively, we define a series of subsets $S_i$ from the entire set of initial seeding integrals:
\begin{align}
S_1 \subset S_2 \subset S_3 \subset \cdots \subset S_{\text{seeding}}\,.
\end{align}
Here, \( S_{\text{seeding}} \) represents the full set of initial seeding integrals, determined through Laporta seeding or improved seeding, such that solving the IBP system generated from the seeding integrals in \( S_{\text{seeding}} \) allows for the solution of all target integrals. By properly defining these subsets, there must exist a subset  \( S_n \) that is the smallest subset sufficient for solving the target integrals.

In general, the memory footprint for IBP reduction is approximately proportional to the size of the IBP system, which in turn is proportional to the number of required seeding integrals. Therefore, our goal is to identify the smallest subset $S_n$. Using seeding integrals from $S_n$ instead of $S_\text{seeding}$ is then expected to reduce memory usage and improve computational efficiency greatly. In other words, $S_n$ is an optimization of $S_{\text{seeding}}$. 

To provide an effective way for finding $S_n$, we define a priority function $F\left(I(n_i);I_{\text{target}}(t_i)\right)$, which calculates the priority of a seeding integral $I(n_i)$ for solving a given target integral $I_{\text{target}}(t_i)$. Here, $n_i$ and $t_i$ denote the indices of the seeding integral and the target integral, respectively. This function is defined on the index space of \( S_{\text{seeding}}  \) and takes values in \( \mathbb{R} \). As an example, in the case of seeding integrals with indices $n_1,\,n_2$, the priority function for Laporta seeding is given by $-n_1-n_2$, which is irrelevant to the indices of the target integral.
With the definition of priority function in hand, the subset of the seeding integrals can be expressed as
\begin{align}
S_i = \{ I(n_i) \mid F\left(I(n_i); I_{\text{target}}(t_i)\right) > f_i \}\,,
\end{align}
where

\begin{align}
f_{\text{max}} \geq f_1 > f_2 > f_3 > \cdots \geq f_{\text{min}}\,.
\end{align}
The upper and lower bounds of $f_i$ are:
\begin{align}
f_{\text{max}} = F(I_\text{target};I_\text{target})\,,
\end{align}

\begin{align}
f_{\text{min}} = \min\big(F(I_\text{any seeding}; I_\text{target})\big)\,,
\end{align}
where $I_{\text{any seeding}}$ represent all initial seeding integrals, and $I_{\text{target}}$ represent the target integrals.

Provided that there exists a sufficiently accurate priority function \( F_0 \) for estimating the priority of the seeding integral $I(n_i)$, a small solvable subset can be defined as:

\begin{align}
\label{eq:smallestSub}
S_{n} = \{ I(n_i) \mid F_0\left(I(n_i); I_{\text{target}}(t_i)\right) > f_c \}\,,
\end{align}
where $f_c$, and consequently $S_n$, can be determined using a binary search algorithm once the corresponding priority function $F_0$ has been identified. Therefore, our goal is to determine the best-estimated priority function $F_0$.

In practical operation, identifying the optimal priority function is a critical challenge. While the optimal priority function may vary for different IBP problems, it is impractical to run the complex IBP reduction thousands of times as an evaluator to test and search for the best-estimated priority function. Considering the scalability and interpretability of FunSearch, we hope to obtain a general best-estimated priority function under a simpler IBP problem and naturally extend it to more complex problems.

Specifically, we perform FunSearch for IBP reductions of the one-loop massless bubble integral, where the corresponding index space of the Feynman integral is two-dimensional. This problem is simple enough while still possessing basic geometric properties. 
The one-loop massless bubble integral is defined as 
\begin{align}
    I(n_1,n_2) = \int \frac{d^d l}{i \pi^{d/2-1}} \frac{1}{(l^2)^{n_1} \big((l-p)^2\big)^{n_2}}\,,
\end{align}
where we set $p^2=1$. The integrals satisfy the following IBP identities:
\begin{align}
0=&\, (d-2n_1-n_2)I(n_1,n_2) - n_2 I(n_1-1,n_2+1) \nonumber \\  
&+ n_2 I(n_1,n_2+1)\,,\\
 0=& \,n_1 I(n_1+1,n_2-1) - n_1 I(n_1+1,n_2) \nonumber \\
&- n_2 I(n_1-1,n_2+1) + n_2 I(n_1,n_2+1) \nonumber \\  
&+ (n_2-n_1)I(n_1,n_2)\,
\end{align}
with zero sectors $I(1,0),\, I(0,1)$ and a single master integral $I(1,1)$.

For this one-loop IBP reduction problem, the priority function can be written as:

\begin{align}
F(I(n_i), I_{\text{target}}(t_i)) = \text{priority}(n_1, n_2; t_1, t_2)\,.
\end{align}
When evaluating the above priority function, we choose $S_{\text{seeding}}$ to be all non-zero integrals inside a square box with a side length of $ 10 + t_1 + t_2 $. This box is large enough to ensure that the final best-estimated priority function is optimized based on the properties of the index space, rather than the choice of the seeding integrals. This allows us to confidently extend the results to more complex IBP reduction problems. For the same reason, the evaluation is performed for several different single-target integrals, and the priority score is determined based on the average size of $S_n$. Under this set of seeding integrals, the Laporta seeding itself also manifests as such a priority:

\begin{align}
\text{priority}_{\text{Laporta}}(n_1, n_2; t_1, t_2) = -n_1 - n_2 \,, 
\label{eq:laporta}
\end{align}

and

\begin{align}
&\nonumber S_{n} = \\&\{ I(n_1, n_2) \mid \text{priority}_{\text{Laporta}}(n_1, n_2; t_1, t_2)\geq -t_1 - t_2 \}\,.
\end{align}

Through the steps outlined above, we can encode the simple one-loop IBP reduction problem into FunSearch using a priority function $\text{priority}(n_1,n_2;t_1,t_2)$. This allows us to find an optimal priority function $\text{priority}_0(n_1,n_2;t_1,t_2)$ via the iterative operation of FunSearch. Subsequently, by identifying the function $F_0(I(n_i);I_{\text{target}}(t_i))$ from $\text{priority}_0(n_1,n_2;t_1,t_2)$, we can naturally extend it to more complex IBP reduction problems, thereby optimizing the IBP reduction in a more general sense. We anticipate that this approach will reduce memory requirements, enhance computational efficiency, and provide a more efficient and cost-effective solution for IBP reduction problems.

\section{Priority function found by FunSearch and generalized by human experts}

Having introduced the concept of the priority function and its application to IBP reductions via FunSearch, we now present the results of optimizing Feynman integral reductions using FunSearch.

Using our method, which combines LLMs, evolutionary algorithms, and a problem specific priority function, we successfully discovered a highly effective priority function for solving one-loop IBP reductions. Compared to the traditional Laporta approach, our method achieves higher computational efficiency and reduced memory usage. Owing to its interpretability and scalability, domain experts can establish an effective feedback loop with the algorithm to easily extend this priority function to more complex, multi-target IBP reductions.

Specifically, in one-loop reduction problem presented above, this priority function reduces the solvable subset $S_n$ by approximately a factor of 5 compared to the Laporta seeding, for the target integral $I(15,10)$. For more complex target integrals, it achieves an even higher improvement factor.  During the operation of FunSearch, several similar best-estimated priority functions repeatedly emerged. An example is shown in Listing.~\ref{lst:priority_function}.

Such best-estimated priority functions cluster around target and master integrals in the index space. The evolution process seems to encourage such a tendency. A representative evolution process is shown in Fig.~\ref{fig:priorities}. Detailed analysis of the evolution process and algorithm execution are presented in the appendix.

\begin{widetext}

\begin{figure}[ht]
    \centering
    \includegraphics[width=\textwidth]{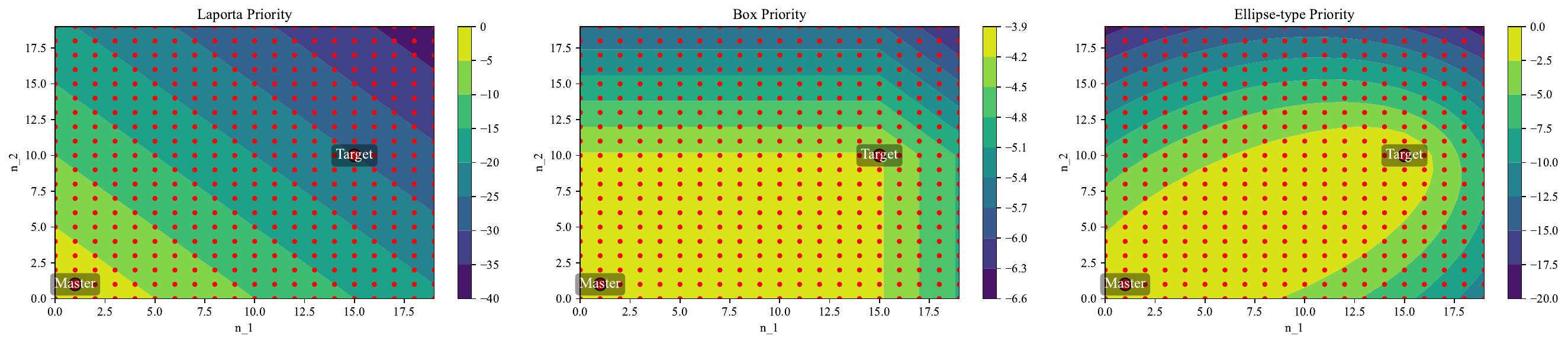}
    \caption{Visualization of the Laporta, Box, and Ellipse-like priority. The target integral $I(15,10)$ is marked as 'Target', and the master Integral is marked as 'Master'.}
    \label{fig:priorities}
\end{figure}
\end{widetext}

\begin{lstlisting}[language=Python, caption={Priority Function example ($m=2$)}, label={lst:priority_function}]
# Code
import numpy as np
def priority(node: tuple[int, int], node_target: tuple[int, int]) -> float:
    x, y = node
    a, b = node_target
    d = ((x - a) ** 2 + (y - b) ** 2) ** 0.5
    p = (x ** 2 + y ** 2) ** 0.5 - (a ** 2 + b ** 2) ** 0.5
    return - (d + p)
\end{lstlisting}

After running FunSearch 10 times, each with 5 000 epochs on one loop IBP reductions problem, we can easily generalize the best-estimated priority function $\text{priority}_0(n_1,n_2;t_1,t_2)$ from the best functions found by FunSearch (for example, Listing~\ref{lst:priority_function}) to:

\begin{align}
\label{eq:BestPriority}
\nonumber&F_0(I(n_i); I_{\text{target}}(t_i)) 
\\&\quad= -\left\{ \left[ \sum_i |n_i|^m \right]^{\frac{1}{m}} + \left[ \sum_i (n_i - t_i)^2 \right]^{\frac{1}{2}} \right\}\,,
\end{align}
where $m$ is a number in set $\{1,2,4,\infty\}$, and the first term is $\text{max}\,(|n_i|)$ for $m=\infty$. The priority function in above equation constitutes one of the main results of this paper and is applicable to Feynman integrals with any number of loops and legs. The second term in the priority function favors seeding integrals without dots or numerators when solving target integrals of the same type, aligning with the objectives of syzygy techniques~\cite{Gluza:2010ws,Larsen:2015ped,Bohm:2017qme,Lee:2014tja,Bitoun:2017nre,Agarwal:2020dye}.

The efficiency of Eq.~\eqref{eq:BestPriority} is evidently scalable. In the next section, we discuss its application to more complex IBP reduction problems and further verify its effectiveness and scalability in large-scale problems.

\section{Applications in complex IBP reductions}
\label{sec:apply}

Building on the success of AI-assistance in optimizing the one-loop IBP reduction problem, we explore its application to more complex, multi-loop Feynman integrals in this section, demonstrating the scalability of our approach. We employ the Ellipse-type priority function given in Eq.~\eqref{eq:BestPriority} with $m=2$. The symbolic IBP identities are generated using \texttt{LiteRed}~\cite{Lee:2012cn,Lee:2013mka}, while \texttt{Mathematica} is employed to construct the initial seeding integrals based on an improved seeding strategy and to sort them according to the best-estimated priority function. Finally, we solve the resulting system of linear equations with \texttt{FiniteFlow}~\cite{Peraro:2019svx}, utilizing its sparse solver capabilities. 

In our implementation of \textit{improved seeding}, we introduce an adjustable vector $(g_0, g_1,\cdots g_n, \cdots )$ to describe the generation process: seeding integrals for sectors with $t= t_p-n$ are produced only if they satisfy $s \leq s_p+ g_n$, starting from a primary sector with $t=t_p$ and $s= s_p$ for the most complicated target integrals. A similar implementation is adopted in Ref.~\cite{Guan:2024byi}. 
The minimal seeding set which enables the solution of target integrals within the improved seeding framework can be determined by searching for the optimized vector $(g_0, g_1,\cdots g_n, \cdots )$ and the parameter $d_\text{max}$, starting from the configuration that generates the fewest seeding integrals. As an example, choosing $g_n=-n$ and $d_\text{max} =0$ could yield a minimal seeding set within this framework. Hereafter, "improved seeding" always refers to the minimal seeding within the improved seeding framework.

The initial seeding integrals generated by the improved seeding strategy can be classified into different categories, which are then ordered based on the values of the priority function. Typically, $\mathcal{O}(10^2)$ categories are produced for $\mathcal{O}(10^5)$ seeding integrals. In most cases, the seeding integrals from only the first few categories are sufficient to reduce the target integrals, significantly lowering memory usage.

To determine the required number of categories, one can directly apply the binary search algorithm to identify the smallest subset $S_n$ as defined in Eq.~\eqref{eq:smallestSub}. However, $S_n$ does not always yield the optimal results in terms of equation-solving time. This can be explained by the fact that the priority function $F_0$ is primarily designed to minimize the number of required seeding integrals, rather than to optimize the time for solving the linear system. Designing a priority function to minimize equation-solving time is left for future work. To provide greater flexibility in selecting a suitable subset and thereby improve both memory efficiency and computational time, we propose the following search algorithm:

\begin{itemize}
    \item[1.] Count the total number of initial seeding integrals, denoted as $N_\text{tot}$.
    \item[2.] From left to right, identify the first $M$ categories in which the seeding integrals collectively span all master integrals. Let the number of seeding integrals in these first $M$ categories be $N_m$.
    \item[3.] Introduce an adjustable relaxation factor $0 \leq a \leq 1$ and define  
    \begin{equation}
    \label{eq:relaxationF}
        N_r = N_\text{tot}^{a} \, N_m^{1-a}\,.
    \end{equation}  
    Identify the first $G$ categories that collectively contain at least $N_r$ seeding integrals. 
    \item[4.] Generate IBP equations separately from the seeding integrals in the first $M$ and first $G$ categories. Solve the resulting linear equations using the sparse solver provided in \texttt{FiniteFlow}. If both cases successfully reduce the target integrals, we stop. If only the second case is sufficient, we use all seeding integrals for reductions and also apply a binary search to refine the selection within the range from the first-$(M+1)$ to the first-$(G-1)$ categories. The optimal choice is determined by the configuration that solves the equations for a numerical sample in the shortest time. If the seeding integrals from the first $G$ categories are still insufficient to reduce the target integrals, we use all available seeding integrals, equivalent to the improved seeding strategy.  
    \item[5.] The final selection of categories directly determines the required seeding integrals.  
\end{itemize} 
The above algorithm also improves the search time compared to a direct binary search approach.

In the following, we present several explicit examples to show the effectiveness of Ellipse-type priority functions. We will first show the examples with single target integrals and then present an example with multiple target integrals. In all examples, we set the relaxation factor $a$ in Eq.~\eqref{eq:relaxationF} to $1/3$. All subsequent computations are performed on the same machine, equipped with a Xeon Gold 6148 CPU and 768 GB of available memory.

\begin{figure}[h]
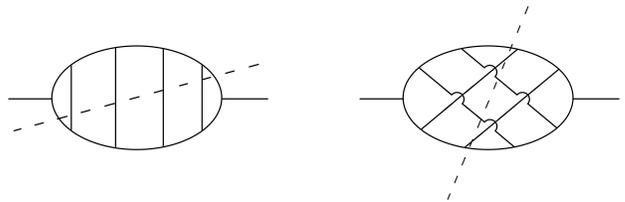

    \centering
    \begin{subfigure}{0.46\linewidth}
        \centering
        \includegraphics[width=\linewidth]{planar.pdf}
        \caption{The planar 5-loop self-energy integral family, where the propagators crossed by dashed lines represent cut propagators.}
        \label{fig:subfig1}
    \end{subfigure}
       \hfill
    \begin{subfigure}{0.46\linewidth}
        \centering
        \includegraphics[width=\linewidth]{non-planar.pdf}
        \caption{A non-planar 5-loop self-energy integral family, where the propagators crossed by dashed lines represent cut propagators.}
        \label{fig:subfig2}
    \end{subfigure}
    \caption{Two six-particle phase-space integral families}
    \label{fig:main}
\end{figure} 

\subsection{A planar six-particle phase-space integral family}

We consider a planar six-particle phase-space integral family, illustrated in Fig.~\ref{fig:subfig1}, which arises in the study of multi-loop scattering amplitudes in quantum field theory~\cite{Ruijl:2017cxj,Gituliar:2018bcr,Magerya:2019cvz,Maheria:2022dsq}. The corresponding propagators are given by
\begin{align}
&D_1 = l_1^2, D_2 = l_2^2, D_3 = l_3^2, D_4 = l_4^2, D_5= l_5^2, \nonumber \\
& D_6 = (l_1-p)^2,  D_7 = (l_2-p)^2, D_8 = (l_3-p)^2,  \nonumber \\
&D_9 = (l_4-p)^2, D_{10}= (l_5-p)^2, D_{11} = (l_1-l_2)^2,  \nonumber \\
&D_{12} = (l_2-l_3)^2, D_{13} = (l_3-l_4)^2, D_{14} = (l_4-l_5)^2, \nonumber \\
& D_{15}= (l_1-l_3)^2, D_{16}= (l_1-l_4)^2, D_{17} = (l_1-l_5)^2, \nonumber \\
& D_{18} = (l_2-l_4)^2, D_{19} = (l_2-l_5)^2, D_{20} = (l_3-l_5)^2 \,,
\label{eq:planar_prop}
\end{align}
where $p$ is the external momentum with $p^2>0$, the propagators $D_1, \, D_{10}, \, D_{11}, \, D_{12}, \, D_{13}, \, D_{14}$ are cut propagators, while the last six propagators are ISPs. This integral family contains 30 IBP identities and 25 master integrals. We choose the following set of integrals to test the effectiveness of priority function $F_0$, the first seven integrals are from the top sector with the first 14 indices being 1: 
\begin{align}
    s_2= &\ (1,\cdots,1, -1,-1,0,0,0,0), \nonumber \\
     s_3= &\ (1,\cdots,1, -1,-1,-1,0,0,0),\nonumber \\
    s_4= &\ (1,\cdots,1, -1,-1,-1,-1,0,0), \nonumber \\ 
     s_5= &\ (1,\cdots,1, -1,-1,-1,-1,-1,0), \nonumber \\
     s_6 = &\ (1,\cdots,1, -1,-1,-1,-1,-1,-1), \nonumber \\
    s_8 = &\ (1,\cdots,1, -2,-2,-1,-1,-1,-1), \nonumber \\
    s_{10} = &\ (1,\cdots,1, -2,-2,-2,-2,-1,-1)\,, \nonumber \\
    s_{12} = &\ (1,\cdots,1, -2,-2,-2,-2,-2,-2)\,,
    \label{eq:integral_s}
\end{align}
and the last three are sub-sector integrals with dots and numerators:
\begin{align}
    d_1 = (1, 0, -1, 1, 0, 1, -1, 1, 0, 1, 2, 1, 1, 1, -2, 0, 0, 0, -1, 0)\,,  \nonumber \\
d_3 = (2, 0, -1, 2, 0, 1, -1, 1, 0, 1, 2, 1, 1, 1, -2, 0, 0, 0, -1, 0)\,, \nonumber  \\ 
d_5 = (2, 0, -1, 2, 0, 1, -1, 2, 0, 2, 2, 1, 1, 1, -2, 0, 0, 0, -1, 0)\,.
\end{align}

Tab.~\ref{tab:sixPlanar} presents a performance comparison of IBP reductions using different seeding strategies: Laporta seeding, improved seeding, and our best-estimated priority function $F_0$, as defined in Eq.\eqref{eq:BestPriority}. For Laporta seeding, we use the C++ package \texttt{FIRE6}~\cite{Smirnov:2019qkx,Smirnov:2023yhb} as a representative package. For improved seeding and seeding from priority function $F_0$, we impose \texttt{FiniteFlow}~\cite{Peraro:2019svx} to solve the sparse linear equation systems. To enhance efficiency of reductions, \texttt{FIRE6} is called with four CPU cores to evaluate a single numerical sample point over finite fields~\cite{vonManteuffel:2014ixa,Peraro:2016wsq}. The number of seeding integrals listed in the second column of Tab.~\ref{tab:sixPlanar} is estimated by dividing the total number of equations generated by \texttt{FIRE6} by the number of symbolic IBP identities. For the planar integral family, the improved seeding pattern $(g_0,g_1,g_2,g_3,\cdots) = (0,-1,-1,-2,\cdots),\, d\leq d_\text{max} =0$ is sufficient and is adopted to reduce the selected integrals.

The results in the table indicate that while \texttt{FIRE6} efficiently manages memory usage, it generates an excessive number of seeding integrals, making it infeasible to compute $s_8$ and beyond within a reasonable time. The improved seeding approach successfully handles $s_8$ but cannot extend beyond that, whereas the priority function $F_0$ enables the computation of integrals up to $s_{12}$ with relative ease, achieving an improvement factor of \textcolor{blue}{24.8} in the number of required seeding integrals. This highlights the crucial role of the priority function method in drastically reducing the computational complexity of IBP reductions, allowing previously intractable integrals to be computed efficiently within practical memory limits. 

\begin{widetext}

\centering
    \begin{table}[ht]
   \centering
    \begin{tabular}{c|c|c|c|c|c}
         Integrals & FIRE6 (4 cores)&  Improved seeding & Priority function $F_0$ & $\frac{N_{\text{I}}}{ N_{\text{P}}}$ & $\frac{T_{\text{I}}}{T_\text{P}}$ \\ 
         \hline
          $s_2$ & 1 299 923/286.7  & 3061/11.8/0.069 & 2985/12.3/0.066  &1.03 &1.05         \\ 
          $s_3$  & 7 276 962/1147.21 & 4125/18.3/0.13                   &     3708/15.4/0.11 & 1.11&1.18     \\ 
          $s_4$ &  20 549 627 /2954.26 &10 803/47.5/0.34                              & 7144/32.9/0.34 & 1.51& 1.0  \\
          $s_5$  & 81 359 547/18779.5  &35 065/165.7/1.5  & 19 052/89/1.25 & 1.84&1.2 \\ 
          $s_6$ &  237 961 845/39245.8 & 108 215/551.9/5.02   & 46 390/226.8/4.0 & 2.33& 1.26\\ 
          $s_8$ &- &791 164/4271.4/36.2 & 101 549/624/20.9 & 7.79& 1.73  \\
          $s_{10}$ & -&4 388 968/OOM &268 909/2198.8/132.7 & 16.32 & -\\
          $s_{12}$ & -&19 901 220/OOM & 802 488/12508.9/1462.3 & \textcolor{blue}{24.8} & - \\ 
          \hline
          $d_1$ &523 596/249 &154 700/686.39/0.07 &1791/8.8/0.06 &86.38 & 1.17  \\  
          $d_3$ &2 940 855/1058.5  & 2 775 780/OOM    &  3466/15.8/0.35 &800.9 & -       \\ 
          $d_5$ & 11 654 385/4523.17  &21 740 796/OOM & 7109/47.7/2.24  & \textcolor{blue}{3058} & -
    \end{tabular}
    \caption{Performance comparison of different IBP reduction methods for selected planar integrals. In column 2, the notation a/b represents (number of seeding integrals) / (time in seconds to solve a single numerical sample over finite fields using \texttt{FIRE6} with four cores). And a dash '-' in this column indicates that the calculation was not performed due to the potentially long runtime, as suggested by the scaling behavior of computation time from $s_2$ to $s_6$. In columns 3–4, a/b/c denotes (number of seeding integrals) / (time in seconds for \texttt{FiniteFlow} to learn the linear system) / (time in seconds for \texttt{FiniteFlow} to solve a single numerical sample over finite fields). The abbreviation 'OOM' indicates an out-of-memory error due to exceeding the 400GB RAM limit. The ratios $\frac{N_{\text{I}}}{N_{\text{P}}}$ and $\frac{T_{\text{I}}}{T_{\text{P}}}$ measure the improvements in the number of seeding integrals and computational efficiency, respectively, when comparing the $F_0$-prioritized approach to the improved seeding method. The priority function $F_0$ exhibits superior scalability, particularly for high-complexity integrals such as $s_{12}$ and $d_5$.}
    \label{tab:sixPlanar}
\end{table}

\end{widetext} 

For subsector integrals with many dots and numerators, the advantage of using the priority function $F_0$ becomes even more pronounced. For example, the integral $d_5$ initially required 21 740 796 seeding integrals, whereas our method reduces this number to just 7109—an improvement by a factor of \textcolor{blue}{3058}. Previously, the IBP reduction process demanded excessive memory, exceeding 400GB RAM, making it impractical for improved seeding strategy. With our approach, the reduction including the reconstruction of $D$ can now be performed on a standard laptop in under two minutes, highlighting the remarkable efficiency and scalability of the priority function $F_0$.

\subsection{A non-planar six-particle phase-space integral family}

In this example, we consider a non-planar six-particle phase-space integral family shown in Fig.~\ref{fig:subfig2}, the propagators take the following explicit form:

\begin{align}
&D_1 = l_1^2, D_2 = l_2^2, D_3 = l_3^2, D_4 = l_4^2, D_5= l_5^2, \nonumber \\
&D_6 = (l_1-p)^2, D_7 = (l_5-p)^2, D_8 = (l_2-l_3+l_5-p)^2, \nonumber \\
&D_9 = (l_1-l_3+l_5-p)^2, D_{10}= (l_1-l_3+l_4-p)^2, \nonumber \\
& D_{11} = (l_1-l_2)^2, D_{12} = (l_2-l_3)^2, D_{13} = (l_3-l_4)^2,\nonumber \\
& D_{14} = (l_4-l_5)^2,  D_{15}= (l_1-l_3)^2, D_{16}= (l_1-l_4)^2, \nonumber \\
& D_{17} = (l_1-l_5)^2, D_{18} = (l_2-l_4)^2, D_{19} = (l_2-l_5)^2,\nonumber \\
& D_{20} = (l_3-l_5)^2 \,.
\label{eq:nonplanar_prop}
\end{align} 
where the propagators $D_3, D_{9}, D_{11}, D_{12}, D_{13}, D_{14}$ are cut propagators, and the last six are ISPs. The integral family contains 30 IBP identities and 46 master integrals. To test the priority function, we take the top sector integrals $s_2$ to $s_6$ which have the same form as shown in Eq.~\eqref{eq:integral_s} and the following integrals with dots and numerators  
\begin{align}
    d_2 = (0, 1, 1, -1, -1, 0, 1, 0, 2, 2, 1, 1, 1, 1, -1, -1, 0, 0, 0, 0)\,,  \nonumber \\
d_3 = (0, 1, 2, -1, -1, 0, 2, 0, 2, 2, 1, 1, 1, 1, -1, -1, 0, 0, 0, 0) \,, \nonumber \\
d_5 = (0, 2, 2, -1, -1, 0, 2, 0, 2, 2, 1, 1, 1, 1, -1, -1, 0, 0, 0, 0)\,.
\end{align}

The non-planar integral family presents significantly greater computational challenges compared to the planar case. The improved seeding strategy used for the planar family is no longer sufficient. Instead, a more refined seeding pattern is required: $(g_0,g_1,g_2,g_3,\cdots) = (0,-1,-2,-3,\cdots),\, d\leq d_\text{max} =1$. 
Tab.~\ref{tab:sixNPlanar} presents a performance comparison of IBP reductions using different seeding strategies: improved seeding, and our best-estimated priority function $F_0$.

For integrals $s_2$ to $s_5$, the improved seeding and priority function $F_0$ yield the same results when using the searching algorithm introduced at the beginning of this section. For the integral $s_6$, the improved seeding strategy is unable to solve the system within the 400GB memory constraint, whereas the priority function $F_0$ successfully handles the reduction. This shows the necessity of using the priority function $F_0$ method even though it could be not quite efficient in time. For sub-sector integrals with many dots and numerators, the priority function $F_0$ demonstrates similar improvements as observed in the planar case. For example, for the integral $d_5$, it results in an improvement factor of \textcolor{blue}{1060}, showcasing the effectiveness of the approach in significantly reducing the number of required seeding integrals.

\begin{widetext}

\begin{table}[ht]
    \centering
    \begin{tabular}{c|c|c|c|c}
         Integrals & Improved seeding & Priority function $F_0$& $\frac{N_{\text{I}}}{ N_{\text{P}}}$ & $\frac{T_{\text{I}}}{T_\text{P}}$ \\
         \hline
          $s_2$ &   34 755/336.28/6.2 &   34 755/322/6.4  & 1 &0.97       \\ 
          $s_3$  & 34 755/360.52/7.54                  &   34 755/320.78/6.34 & 1 &1.19        \\ 
          $s_4$  &59 157/1383/26.2                              &  59 157/1383/25.5 &1 & 1.03 \\
          $s_5$  & 160 377/10361.7/151.7 &  160 377/10300.45/155.1   & 1 & 0.98 \\ 
          $s_6$ & 492 401/OOM   & 93 297/15626.5/2539.3 & {5.3} &- \\ 
          \hline
          $d_2$ & 598 428/7956.6/5.7  &1337/11.6/0.58 & 447.59 & 9.83\\  
          $d_4$ & 4 779 738/OOM &4709/56.4/5.05 & 1015.02 & -          \\ 
          $d_5$ &  11 519 163/OOM &  10 859/207.8/16.2  & \textcolor{blue}{1060} & -
    \end{tabular}
    \caption{Performance comparison of IBP reductions using different seeding patterns for selected non-planar integrals. In this table, we do not include results for Laporta seeding. The format follows the same conventions as those in Tab.~\ref{tab:sixPlanar}.}
    \label{tab:sixNPlanar}
\end{table}

\end{widetext}

\subsection{An example with multiple target integrals}
In the above two examples, the priority function is applied to the cases with only a single target integral. In this example, we apply priority function to the case with multiple target integrals. For simplicity, we apply the priority function recursively to the target integrals, each time we only choose the most complicated unsolved target integral. Since we are able to solve a large fraction of target integrals each time, applying the priority function recursively a few times can give solutions for all target integrals.

We take the integral family from the first example, and derive dimensional recurrence relation~\cite{Tarasov:1996br} for the top sector integral $s_1=(1,\cdots,1,-1, 0,0,0,0,0)$ by expressing it in $D+2$ dimension in terms of linear combinations of integrals in $D$ dimension. This procedure generate 955 target integrals, among them the most complicated integrals have $t=14,\, s=6, \, d=0$. For this example, we find that applying the priority function one time is able to derive the above recurrence relation. The priority function method generates 129 333 seeding integrals, and takes \texttt{FiniteFlow} 948.4 seconds to learn the linear system and 13.3 seconds to solve the system on one numerical sample point. As a comparison, the improved seeding method with seeding pattern $(g_0,g_1,g_2,g_3,\cdots) = (0,-1,-1,-2,\cdots),\, d\leq d_\text{max} =0$ generate 255 929 seeding integrals, and takes \texttt{FiniteFlow} 2733.56 seconds and 14.1 seconds to learn and solve the system on one numerical sample point, respectively. This again demonstrates the improvement of the priority function method over the improved seeding method.

\section{Conclusions}

In this work, we introduced a novel approach to optimizing the reduction of Feynman integrals using integration-by-parts (IBP) identities. By developing a priority function method through the FunSearch algorithm—a combination of large language models (LLMs) and genetic algorithms—we have identified the best-estimated priority function $F_0$ as shown in Eq.~\eqref{eq:BestPriority}. The method has led to significant improvements in both memory usage and computational efficiency compared to traditional methods, such as the Laporta seeding and improved seeding strategies. 

Our best-estimated priority function $F_0$ applies to, and has been rigorously tested on, a diverse set of Feynman integrals, including both single-loop and multi-loop cases with planar and non-planar scenarios. Specifically, we have examined planar and non-planar six-particle phase-space integral families in multi-loop scenarios. The results demonstrate that our method can substantially reduce required seeding integrals, leading to significant memory savings and faster computation times. For instance, for top sector integrals in the case of the planar six-particle phase-space integral family, we observed a reduction in the number of seeding integrals by up to a factor of 24.8, whereas in the non-planar case, the reduction reached a factor of 5.3. For subsector integrals with many dots and numerators, the advantage of using the priority function $F_0$ becomes even more pronounced, achieving an improvement factor of 3058 in the planar case and 1060 in the non-planar case. More importantly, we observed that as the complexity of the Feynman integrals increases, the improvement factor achieved grows larger. These findings underscore the effectiveness and scalability of our approach, especially for integrals of high complexity.

The success of our method is attributed to the unique synergy between large language models (LLMs) and evolutionary algorithms, which enables the generation of efficient and interpretable priority functions. This framework provides a more optimized solution for IBP reductions and offers a scalable and generalizable approach that can be extended to more complex problems. Moreover, the interpretability of the priority functions generated by FunSearch allows domain experts to further refine and adapt these functions for specific applications, thereby enhancing their practical utility.

In summary, our work contributes to applying AI methods to high-energy theoretical physics by proposing a new tool for optimizing the reduction of Feynman integrals. Combining the FunSearch algorithm with a priority function offers a potentially more efficient and scalable framework for addressing some of the computational challenges associated with multi-loop calculations. This development could help facilitate more practical evaluations of scattering amplitudes and cross-sections, supporting researchers in studying higher-order processes and complex topologies in quantum field theory. We hope this work inspires future efforts to optimize priority functions and explore further applications of explainable AI methods in theoretical physics.

\section*{Acknowledgements}
We would like to thank T. Peraro for the helpful instructions on the usage of \texttt{FiniteFlow}. T.-Z.Y. is supported by the European Research Council (ERC) under the European Union's Horizon 2020 research and innovation programme grant agreement 101019620 (ERC Advanced Grant TOPUP). H.X.Z. is supported by the National Science Foudantion of China under contract No. 12425505 and Asian Young Scientist Fellowship. The work of Q.-H. C. is partly supported by the National Science Foundation of China under Grant Nos. 12235001. M.-X.L. is supported by National Natural Science Foundation of China under contract No. U2230402.

\textbf{Note Added}: While finalizing this paper, we became aware of Ref.~\cite{funIBP}, where FunSearch was applied to IBP from a different perspective. In Ref.~\cite{funIBP}, FunSearch is used to search for conditional expressions, whereas in our case, it is employed to search for a mathematical priority function.

\bibliographystyle{apsrev4-1}
\bibliography{main}

\appendix

\section{FunSearch Operation and Hyperparameter Selection}

To provide further insight into the implementation and optimization of FunSearch, we include this appendix detailing the selection of hyperparameters and an analysis of the evolution process. This information complements the main results presented in the preceding sections.

\subsection{Hyperparameter Selection}

The selection of hyperparameters significantly impacts the performance and outcomes of FunSearch. Below lists some key hyperparameters and their selection criteria:

\textbf{Model:} The CPM-2B model~\cite{minicpm2024} was selected due to its superior performance in code generation tasks. It does not require additional fine-tuning and has shorter invocation times compared to other models such as Qwen-14B~\cite{qwen}, which showed similar performance but with longer processing times.

\textbf{Temperature:} The temperature of the model was set within the range of 0.9 to 1.1. Temperatures above 1.2 led to non-functional generated functions, while temperatures below 0.9 resulted in excessive repetition in model outputs. At a temperature of 0.01, the output efficiency decreased by nearly 10 times. As shown in Fig.~\ref{fig:funsearch}

\textbf{Islands:} The number of islands was set to 10. This value was chosen based on literature recommendations~\cite{funsearch} and effectively balances computational resources and exploration efficiency.

\textbf{Island Temperature:} The island temperature was set within the range of 10 to 100. This parameter affects the diversity of programs within an island. Higher temperatures help prevent the algorithm from falling into local optima.

\textbf{Cleanup Frequency:} The cleanup frequency was set to 50 to 100 epoch per cleanup. This parameter ensures that the worst-performing islands are regularly cleaned up, maintaining population diversity.
After each cleanup, the best code on the best and second-best islands are placed back into the cleaned islands. This ensures that the population maintains a high level of quality.

\textbf{Evolution Rounds:} The number of evolution epochs was set between 5000 and 100000. An inDistributionitial value of 5000 rounds is reasonable and can be adjusted.

\textbf{Output Length:} The output length matters and the output length was set within the range of about 1024 to 4096. Short output lengths result in overly simple programs that fail to effectively solve problems, while long output lengths increase the risk of overfitting due to excessive diversity. A range of 2048 to 4096 is considered reasonable.

We allowed priority flipping so the efficiency of FunSearch can increase. That is, If priorities outside Laporta seeding are generally higher than those inside, adding a sign does not affect the function's shape and ensures that priorities expand from low to high complexity, maintaining scalability.

The Laporta priority function in Eq.~\eqref{eq:laporta} was used for initializing the islands. This function is the most widely used and ensures that priority evolution does not start from a specific local minimum. Empirically, evolution starting from Laporta priority often converges fastest. Other designed functions tend to overfit or converge to lower-scoring priorities.

These hyperparameter settings and initialization were carefully chosen to optimize the performance of FunSearch while maintaining computational efficiency and solution quality.

\subsection{Brief Analysis of the Evolution Process}
In the experiments, we executed FunSearch for one-loop IBP reductions a total of 10 times, with each run comprising 5000 epochs. The initial seeding employed was the Laporta priority~\ref{Laporta}. During each epoch, the priority function was evaluated based on its average seeding integrals for 30 distinct fixed target integrals in the top sector of the one-loop IBP reductions. Specifically, the scoring mechanism was defined as follows:
\begin{align}
\text{Score} = 200 - \frac{\sum_{j=1}^{30} S_j}{30} - \frac{L}{400},
\end{align}
where $S_j$ denotes the minimal number of seeding integrals required to solve the j-th target, and $L$ represents the length of the text of the priority function.Here the second term was introduced to penalize overly complex priority functions, thereby encouraging more interpretable forms.

   The best estimated priority function discovered is given in Eq.~\eqref{eq:BestPriority}. In single-loop reduction, the size of the small set $S_n$ corresponding to this priority is about 5 times smaller than that of the Laporta seeding. Out of 10 runs, 7 converged to the best-estimated priority function.

   During the evolution, there are always 3 platforms, namely Laporta priority (average seeding integral number is about 200), Box priority (average seeding integral number is about 100), and Ellipse-type priority (average seeding integral number is about 50). A set of typical function examples is as follows:

   - \textbf{Laporta Priority}: The seeding integral number is approximately $(\Sigma n_i)^D$, where $D$ is the dimension of the indices in IBP reductions.
   
   - \textbf{Box Priority}: The seeding integral number is approximately $\Pi n_i$.
   
   - \textbf{Ellipse-type Priority}: The seeding integral number is approximately $b \times \Sigma n_i$, where $b$ is the width of the ellipse corresponding to the small set $S_n$.

\begin{figure}[htbp]
       \centering
       \includegraphics[width=0.5\textwidth]{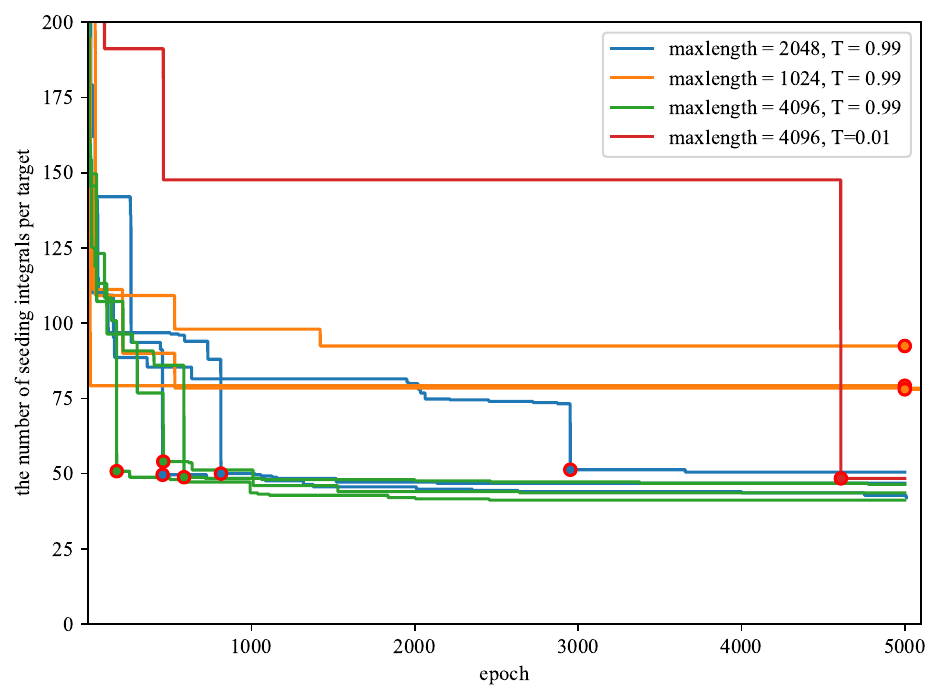}
       \caption{Evolution of the average seeding integral number with different hyperparameters. Each line represents a different experiment. Red dot for each line represents the best priority function achieved in this experiment.}
       \label{fig:funsearch}
   \end{figure}

   The changes in the average seeding integral number of the best estimated priority during the evolution process are shown in Fig.~\ref{fig:funsearch}.

   \begin{lstlisting}[language=Python, caption={Laporta Priority},label={Laporta}]
   # Epoch 9: Score = 152.2
   # Code
   import numpy as np
   def priority(node: tuple[int, int], node_target: tuple[int, int]) -> float:
       return -np.sum(node)
   \end{lstlisting}

   \begin{lstlisting}[language=Python, caption={Box Priority},label={Box}]
   # Epoch 400: Score = 177.5
   # Code
   import numpy as np
   def priority(node: tuple[int, int], node_target: tuple[int, int]) -> float:
       n0, n1 = node
       n_target0, n_target1 = node_target
       d = 4
       dist0 = max(abs(n0), abs(n_target0))
       dist1 = max(abs(n1), abs(n_target1))
       return - (dist0 + dist1) / (d + 2)
   \end{lstlisting}

   \begin{lstlisting}[language=Python, caption={Ellipse-type Priority},label={Ellipse}]
   # Epoch 4775: Score = 188.4
   # Code
   import numpy as np
   def priority(node: tuple[int, int], node_target: tuple[int, int]) -> float:
       x, y = node
       a, b = node_target
       d = ((x - a) ** 2 + (y - b) ** 2) ** 0.5
       p = (x ** 2 + y ** 2) ** 0.5 - (a ** 2 + b ** 2) ** 0.5
       return - (d + p)
   \end{lstlisting}

\textbf{Figure~\ref{fig:funsearch}} demonstrates how the average seeding integral number of the best estimated priority evolves over the number of epochs under different hyperparameters. As the \texttt{maxlength} increases, the priority function evolves to allow more intermediate variables. On the other hand, the penalty term ensures that we always strive to find shorter and more interpretable priority functions. Consequently, as the length increases from 1024 to 2048, the final converged result transitions from the Box priority to the Ellipse-type priority.
   
Moreover, the temperature parameter significantly affects the diversity of FunSearch. Lower temperatures slow down the convergence, but as long as the \texttt{maxlength} permits, the algorithm eventually converges to the Ellipse-type priority. This highlights the importance of balancing exploration and exploitation in the evolutionary process.

Initialization also plays a crucial role. In our experiments, only the Laporta priority as the initial seeding allowed the algorithm to escape local minima and provide effective heuristic priority functions. Other initializations tended to get trapped in local minima, failing to offer any meaningful improvements.

\section{Executing FunSearch on Complex IBP Reductions}
We attempted to apply FunSearch to more complex IBP reductions problems, specifically focusing on two-loop IBP reductions. Despite running the algorithm for approximately 2 days of CPU time and 2000 epoches, no results surpassing the Box priority were observed. This suggests that the complexity of the problem space significantly increases with the addition of complexity, making it more challenging for the algorithm to find improved priority functions within a reasonable timeframe.

Additionally, we explored the performance of FunSearch in scenarios involving multiple target points. Over 30000 epochs (20 islands, CPU time 10879 mins), the best estimated priority obtained is shown in Listing~\ref{mutinode}.

\begin{lstlisting}[language=Python, caption={Best Approximation Priority for Multi-Target},label={mutinode}]
# Epoch 18834: Score = 153.4
# Code
import numpy as np
def priority(node: tuple[int, int], *node_targets: tuple[int, int]) -> float:
    return -max((np.linalg.norm([x - node[0], y - node[1]]) ** 2 + node[0] ** 2 + node[1] ** 2 for x, y in node_targets))
\end{lstlisting}

Its expanded form is given by:

\begin{align}
F_0(I(n_i);& I_{\text{target}j}(t_{ji})) \nonumber \\
&= -{\left[ \sum n_i^2 + \max_j{\left( \sum_i (n_i - t_{ji})^2 \right)} \right]}\,.
\end{align}

This result indicates that even with multiple target points, the algorithm was able to identify a priority function that is at least on par with the Ellipse-type priority, which is the best-estimated priority function we have observed. Testing its full performance in all scenarios is left for future work. This suggests that the algorithm has the potential to find effective priority functions in more complex IBP reductions scenarios, but further evaluation is needed to confirm its superiority.


\end{document}